\let\va=\varphi
\def\pri{^\prime}
\let\la=\lambda

\let\al=\alpha

\let\ra=\rightarrow
\let\va=\varphi

\let\vare=\varepsilon
\let\la=\lambda
\let\vfi=\varphi

\let\ovl=\overline

\let\ti=\times
\let\l=\ldots

\let\ovl=\overline

\let\disp=\displaystyle
\def\que#1#2{\displaystyle\frac{#1}{#2}}

\documentclass[pra,twocolumn]{revtex4}

\usepackage{amsfonts,amssymb,bm,graphicx}

\begin{document}

\title{Suggestions on Photons and on Fermions}

\author{R. Alvargonz\'alez and L. S. Soto}

\affiliation{Facultad de F\'{\i}sica,  Universidad Complutense,
28040 Madrid, Spain}

\date{\today}

\begin{abstract}

In this paper we suggest a configuration of photons consistent with a spin
$\hbar$, and a configuration of the fermions coherent with a spin $\hbar/2$.
These suggested configurations open the way to further analyses which lead to
the following conclusions:
\begin{itemize}\itemsep=0pt
\item There cannot exist elementary particles of spin $\hbar/2$ with a mass
inferior to $1m_e$ or with a radius greater than $1l_e$.
\item The elec\-tros\-ta\-tic force derives from the centrifugal forces inherent to
the spin and are propagated by photons.
\item The derivation of the elec\-tros\-ta\-tic force explains the existence of
positive and negative charges and Coulomb's law.
\item The enormous differences between the centrifugal forces and the
centripetal forces at the surface of the protons give rise to quantic
fluctuations of space which generate the energy flows necessary for
equilibrium. These energy flows can explain gravitation and the strong force.
\item  The mass of the proton, $m_p$, and the mass of the neutron, $m_n$,
must each have a concrete value required for the cohesion of the atomic
nuclei. The suggested hypotheses show that the relation $m_n/m_p$ must be very
slightly greater than 1.00, and that, by a first approximation, both $m_n$ and
$m_p$ must be slightly less than 1851 $m_e$.
\item There are no ``gravitons'' and no ``gluons''; the ``messenger particles''
are always the very real photons.
\end{itemize}
\end{abstract}

\maketitle

\section{Preamble}

Goudsmit and Uhlenbeck discovered electron spin in 1925, while trying to
find out what causes very close parallel lines to appear in the optical
spectrum of hydrogen. They were able to show that these lines could be due to
the fact that the electron possesses a natural angular impulse and bipolar
magnetic momentum, whose component must be equal to $\hbar/2$ or $-\hbar/2$.

When describing the circumstances of the discovery, Uhlenbeck commented that
both he and Goudsmit  imagined that the electron would be configured as a sphere
in permanent rotation, but that they realised that, given the mass of the
electron a spin momentum of $\hbar/2$ would require the tangential velocity
at its equator to be many times the speed of light. Two years later, the
experiments of Phipps and Taylor showed clearly that the electron does in fact
possess a natural angular impulse, $s$, whose value is $\hbar/2$, and soon
after, Dirac used Schr\"odinger's equation while substituting the classic
equation $E=(p^2/2m_0)+V$ by its relativistic form
$E=(c^2p^2+m_0c^4)^{1/2}+V$, to show that the electron must have a spin of
$s=\hbar/2$. Robert Eisberg hailed Dirac's achievement as ``a great victory
for the theory of relativity, since it established electron spin on very sound
theoretical foundations, and showed that it is intimately connected to
relativity". [1] pp. 326-27.

In 1916 Nernst suggested that the quantic fluctuations of space must cause an
electromagnetic radiation which would therefore be inherent to space and,
consequently, have a relativistically invariant spectrum.

In 1958 Sparnaay found this radiation when he was measuring the Casimir effect
at temperatures close to absolute zero. He detected some radiation, which was
independent of temperature and whose spectrum was such that the intensities
of its flows are inversely proportional to the cubes of the wavelengths, which
is a necessary condition for the radiation to be relativistically invariant.
In 1997, S.K. Lamoreaux carried out new measurements of the intensity of the
energy flow of zero-point radiation, using a different method, and reached the
same measurements as Sparnaay's.

A function of spectral distribution which is inversely proportional to the
cubes of the wavelengths implies a distribution of energies which is inversely
proportional to the $4^{\rm th}$ power of the wavelengths, because the
energies of the photons are inversely proportional to their wavelengths. in
1969, Timothy H. Boyer showed that the spectral density function of zero-point
radiation is:
$$f_{\va}(\la)=\que1{2\pi^2}\que1{(\la_\ast)^3},$$
where $\la_\ast$ is the number giving the measurement of the wavelength $\la$.

This function produces the next for the corresponding energies.
$$E_\va(\la)=\que{1}{2\pi^2}\que{hc}{\la} \que{1}{(\la_\ast)^3}$$

For $\la\ra0$ \ $E_\va(\la)\ra\infty$. There must be, therefore a threshold
for $\la$, which hereafter will be designated by the symbol $q_\la$.

\vskip 6pt
\begin{center}*\ \ \ *\ \ \ *\end{center}\vskip 16pt

This paper is closely related to ``Interactions between zero-point radiation
and electrons" ([2] in REFERENCES) and we include here the following extracts
from it, to facilitate the subsequent argument.

Zero-point radiation proceeds equally from all directions of space, and its
interactions with electrons could play the role of the ``Poincar\`e tensions",
preventing the electrons from shattering as a result of the repulsion of their
charge against itself. For this to be the case, there must operate the
equation:\vspace*{-6pt}
$$x^3=\que{4\pi^3}{3\al}(k_\la)^4(r_x)^4[B]_m;\eqno{(1)}\vspace{-6pt}$$
equation (17) in [2], where:
\begin{description}\itemsep=-1pt
\item[$x=$] measurement of the wavelength of the photons with the greatest
energy in zero-point radiation, expressed in $q_\la$ (quanta of wavelength).
\item[$k_\la=$] measurement of the unit of length, $l_e$, in the $(e,m_e,c)$
system, expressed in $q_\la$.
\item[$r_x=$] measurement of the radius of the electron, expressed in $l_e$.
\end{description}
$$[B]_m=\que7{48}B-\que{11}{50}B^2+\cdots+T_mB^m;$$
where $B=\que{2\pi}\al\left(\que{k_\la}\al\right)$ and\vspace*{-6pt}
$$T_m=(-1)^{m-1}\left[\que1{m+1}+\que2{m+2}-\que3{m+3}-1-\que{m(m-1)}6\right]
\que1{m+3},\vspace*{-9pt}$$
p.9 in [2].

The hypothesis that zero-point radiation is also the effective cause of
gra\-vi\-ta\-tio\-nal attraction between two electrons leads to the equation:
$$x^3=\que{2\pi^3}{3\al}\que{(k_\la)^2(r_x)^2[B]_m}{G_e};\eqno{(2)}$$
equation (20) in [2].

In [2] it was also deduced that
$$r_x=1l_e$$
$$k_\la=\left(\que1{2\pi G_e}\right)^{1/2},\qquad
G_e=\que1{2\pi(k_\la)^2}\eqno{(3)}$$
where $G_e$ is the numerical coefficient of the gravitational constant in the
$(e,m_e,c)$ system $q_\la=\al^{1/2}L_P$, where $L_P$ is the Planck length.
$$\left.\begin{array}{l}
k_\la=8.143375\ti10^{20}\\[+8pt]
x=5.257601\ti10^{27}\end{array}\right\}\eqno{(4)}$$

This paper uses the $(e,m_e,c)$ system of units in which the basic
magnitudes are $=$ electron charge, $e$, electron mass, $m_e$, and the speed
of light, $c$. The unit of length is the classic radius of the electron
$l_e=e^2m_e^{-1}c^{-2}$, the unit of time is $t_e=e^2m_e^{-1}c^{-3}$, and that
of quantity of action is $h_e=e^2c^{-1}$. Therefore $h=\que{2\pi h_e}\al$, but
he symbol $h_e$ will not be used.

\section{Conjectures on Photons}

We know that photons have the following properties:
\def\labelenumi{\alph{enumi})}
\begin{enumerate}\itemsep=-2pt
\item They are ``energy packets"\ which possess the quantity of action
$h=\que{2\pi}\al\,\que{m_el_e^2}{t_e}$,
and can behave as particles.
\item They move at the speed of light along rectilinear trajectories. These
are geodesic lines, in a space without curvature. Within the space defined by
the 3-dimensional spherical surface $w^2+x^2+y^2+z^2=R^2$,
the geodesic lines are great circles, but in the case of our Universe, the
immense size of $R$ makes it very difficult not to confound them with straight
lines.
\item Their wave-motion follows planes perpendicular to their trajectories,
producing a spin of $s=\hbar=\que{m_el_ec}\al$.
\item They only differ from each other in the quantity of energy, $E$, which
is contained in the packets in question. This quantity is given by the
equations $E=h\nu=hc/\la$, in which $\nu$ is the frequency (number of photons
per time-unit) and $\la$ is the wavelength. Since $\nu$ and $\la$ are linked
by the relation $\la\nu=c$, we need to consider only one of these variables.
Our aim, to investigate possible configurations of photons, implies that it is
preferable to characterise them according to their wavelength $\la$.
All  photons of wavelength $\la$ are identical.
\end{enumerate}

We must make a brief digression here. The wave-motion of photons is very
diffe\-rent from that of the waves of the sea. Except where they break, the
wave-motion of the sea waves is produced as a vertical movement, and is
propagated through a medium which does not move except in that direction. This
wave-motion does not carry with it anything of substance, being merely a
wave-motion within a uniform substance. In the case of photons, their
wave-motion does carry substance with it, that of the ``energy packets"\
having a quantity of action $\hbar c/\la$  which is what
they are, while they are not
limited in their movement to the wave-motion which gives rise to their spin
$\hbar$, but advance in a straight line at the speed of light.

We will now analyse the following suggested configuration for photons:

The substance of the photons undulates along cylindrical helices, in such
a way that the advance of one wavelength along its rectilinear trajectory is
accompanied by an advance of an equal distance over a circumference of radius
$R=\la/2\pi$, having its centre on the said trajectory, and situated on a
plane perpendicular to it, so that for every complete rotation over that
circumference, there will be an advance of one wavelength over the rectilinear
trajectory.\vskip 6pt

\begin{center}*\ \ \ *\ \ \ *\end{center}\vskip 16pt

To help the analysis of the suggested configuration, there are shown in
Fig. 1 the projections over the planes $ZOX$ and $ZOY$ of a cylindrical
helix, whose equations are:
$$\left.\begin{array}{l}
x=ct\\[+4pt]
z=R\sin\left(\que xR\right)\\ \\[-8pt]
y=R\cos\left(\que xR\right)\end{array}\right\}\eqno{(5)}$$

\begin{figure}[h]
\centering
\resizebox{0.70\columnwidth}{!}{\includegraphics{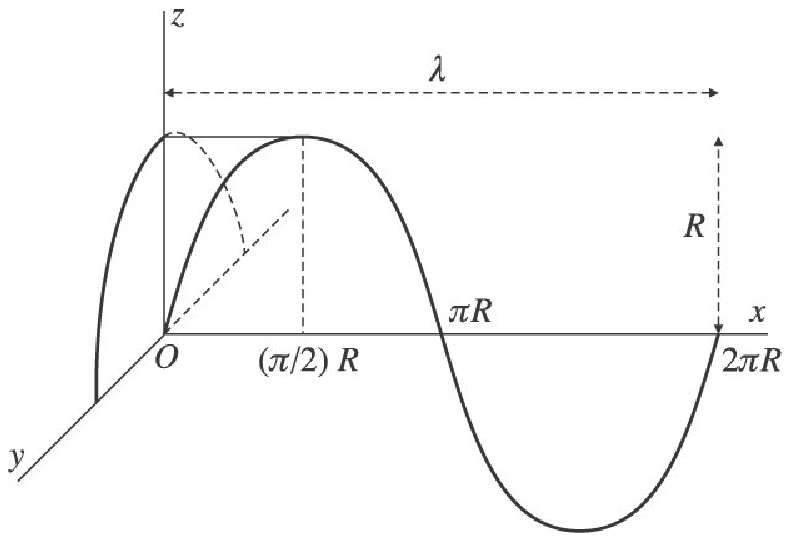}}
\caption{Fig. 1}
\end{figure}

During the time elapsed between $t=0$ and $t=2\pi Rt_e$, the point on the
helix has advanced a distance of $2\pi Rl_e$ along the axis $\ovl{OX}$, and
has made one complete rotation around that axis, which presupposes a journey
of $2\pi Rl_e$ over the circumference of the projection
on the plane $ZOY$. Both during this journey and during that over
$\ovl{OX}$, the velocity has been $c$ (obviously, these speeds are not to be
added together, but seen as independent of each other).

The advance of $2\pi R$ over the axis $\ovl{OX}$, i.e. one wavelength, $\la$, is
matched by a length of the helix traced over the cylinder of rotation with a
radius of $Rl_e$, which is given by
$$L=4R\int^{\pi/2}_0(dx^2+dy^2+dz^2)^{1/2}=$$
$$=4R\int^{\pi/2}_0\left(1+\cos^2\que xR+\sin^2\que xR\right)^{1/2}dx=
(2\sqrt{2})\pi R$$

The length of the cylindrical helix corresponding to a distance of one
wavelength is equal to $\sqrt{2}\la$.

Seeing that electromagnetic wave-motion must run over a cylindrical rotational
surface, it must do so following a cylindrical helix, since for cylinders,
whether rotational or not, their helices are their geodesic lines. For every
distance of one wavelength, $\la$, along $\ovl{OX}$, there can be only one
corresponding photon.

A photon whose energy is equal to $m_ec^2$ (the energy equivalent of the mass
of the electron) has a wavelength of $\la_e=\que{2\pi l_e}\al$, to which there
corresponds $R=l_e/\al$, and its spin momentum will be
$$S=R m_e  c=\que{m_el_ec}\al=\hbar,$$
which is the right value for photons.

\section{Conjectures on the Fermions}

In order to develop the hypothesis proposed in this paper, it will be better
to limit ourselves initially to considering the electron, and then move on to
the remaining elementary particles. The basic suppositions for all of them are
the following:
\def\labelenumi{\arabic{enumi})}\begin{enumerate}\itemsep=-2pt
\item The transformation of a photon into a particle does not require any
change in its substance, but only a change in the organisation of that same
substance. It continues to be the same ``packet of energy with $h$ quantity of
action"\ which, instead of moving lineally and forming part of a train of
waves moving at the speed of light, has turned round on itself to ``bite its
tail", creating an isolated entity rotating in such a way that at any point on
it, the tangential velocity is $c$.
\item The photon does not lose its essentially linear nature in favour of a
spherical surface. What happens is that it rolls itself around a sphere of
radius $R$, in such a way that during the period $Rt_e$ it makes the journey
of $AB=Rl_e$ around the great circle $\phi$ of radius $R$, which meanwhile
completes a rotation of $\psi=Rl_e$ radians around the axis $\ovl{OZ}$
(see Fig. 2). At the end of the said period $Rt_e$, the point which was
initially  at $A$  has moved to $N$, whose coordinates are:
$$\begin{array}{l}
x=R-ON\cos\va\cos\psi\\[+4pt]
y=ON\cos\va\sin\psi\\[+4pt]
z=ON\sin\va\end{array}$$

And where $\va=\psi$; \ $ON=R$:
$$\left.\begin{array}{l}
x=R(1-\cos^2\va)=R\sin^2\va\\[+4pt]
y=R\cos\va\sin\va\\[+4pt]
z=R\sin\va\end{array}\right\}\eqno{(6)}$$\end{enumerate}

\begin{figure}[h]
\centering
\resizebox{0.70\columnwidth}{!}{\includegraphics{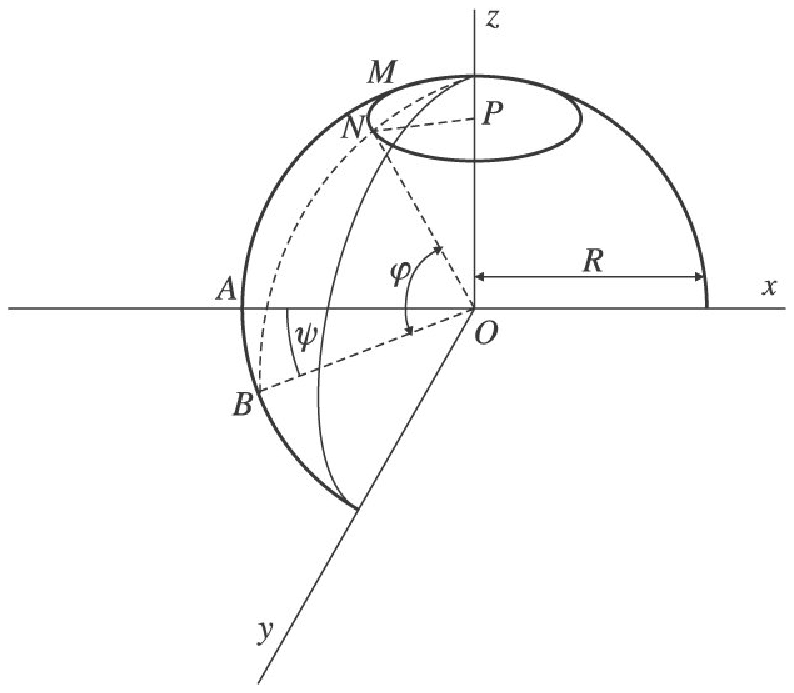}}
\caption{Fig. 2}
\end{figure}

The journey of the photon over curve (6) may be considered
as being made over a great circle
which turns around the axis $\overline{OZ}$ in such a way that the angles
$\vfi$ and $\psi$ are equal. The journey over the great circle is equivalent
to the photon's journey over $\overline{OX}$ in Fig. 1. Both have a length of
$2\pi R$, and neither of them generates angular momentum. During its journey
over the great circle, the angular momentum would follow a diameter at the
equator. However when it has completed a rotation around every point on the
equator, it will be balanced by its opposite going in a contrary direction so
that the sum of the two is zero.

The rotation around $\overline{OZ}$ generates at every point a momentum
determined by the tangential velocity $c\overline{NP}/R=c\cos\vfi$ and by
the radius $NP$. In other words:
$$\hbox{Dif. spin}=\que{m_ed\vfi}{2\pi} c\cos\vfi R\cos\va=
\que{m_eRc}{2\pi}\cos^2\va d\va,$$
where the value $\que{m_ed\va}{2\pi}$ is determined by the equal distribution
of the energy of the photon in a journey of $2\pi$ radians. Therefore we
arrive at:
$$s=\que{m_ecR}{2\pi}4\int^{\pi/2}_0\cos^2\va d\va=\que{m_ecR}2.$$
For $R=\que{l_e}\al$,\ \ $s=\que{m_el_ec}{2\al}=\que\hbar2$.

From equations (6) we obtain:
$$\begin{array}{l}
dx=2R\sin\va\cos\va\,d\va=R\sin2\va\,d\va\\[+4pt]
dy=R(\cos^2\va-\sin^2\va)d\va=R\cos2\va\,d\va\\[+4pt]
dz=R\cos\va\,d\va\end{array}$$

Therefore, $ds=R(\sin^22\va+\cos^22\va+\cos^2\va)^{1/2}d\va=
R(1+\cos^2\va)^{1/2}d\va$, whence
$ds=R\sqrt2\left(1 - \que12\sin^2\va\right)^{1/2}d\va ;$
and, finally:
\begin{eqnarray*}
\frac{1}{4} s & = & R \sqrt{2}\int^{\pi/2}_0
\left(1- \que12 \sin^2\va \right)^{1/2}d\va \\
& = &\sqrt{2} \, R\, 1.3506 = 1.91004 R,
\end{eqnarray*}
which is somewhat greater than the distance travelled along the great circle
between $\va=0$ and $\va=\pi/2$. The length which corresponds to the wave
motion is $1.91R$ compared with $(\pi/2)R=1.57R$, which means a relation
1.216. In Fig. 1 the relation was $\sqrt{2}=1.414.$

For a photon with energy $m_ec^2$, we obtain $\la_e=2\pi l_e/\al$,
$R=l_e/\al$. The radius of the electron is $l_e$ and its circumference $2\pi
l_e$. Therefore a ``rolling up'' of $\la_e$ would mean making
$\que1\al=137.0360$ turns around a circumference
of radius $l_e$, i.e. a
length of $(2\pi/\al)l_e$, which would be covered in a period of
$(2\pi/\al)t_e$ if a speed of $c$ is maintained. We do not need to introduce
a new diagram or new equations; but need only imagine Fig. 2 at a scale of
$\al$, i.e. 1/137.0360 and with the wavelength of the photon, $\la_e$, rolled
up 137.0360 times around the sphere whose radius will now be
$l_e$. In equations (6), it is sufficient to replace $R=l_e/\al$ by $r_e=1l_e$.
The energy $m_ec^2$ will be distributed equally over the length of the
wavelength $\la_e=(2\pi/\al)l_e$,
so that the value of the spin momentum will still be $\hbar/2$.
If at any time the ``rolling-up'' should come undone while the ``tail-biting''
remains, centrifugal force would cause it to change into a circumference of
radius $R_e=l_e/\al$, rotating at the same tangential velocity $c$ and
generating the angular momentum $\hbar$.

A great circle of radius $l_e$, making one turn around the axis $\ovl{OZ}$
every $2\pi t_e$, i.e. every $5.90506\ti10^{-23}$ seconds, would be very hard
to distinguish from a spherical surface of radius $l_e$. As noted above,
Goudsmit and Uhlenbeck discovered in 1925 that the electron must have a spin
of $\hbar/2$, and deduced that if it were configured as a spherical surface,
the radius of that surface would have to be greater than that of an entire
hydrogen atom; otherwise, the tangential velocity at its equator would have to
be much greater than that of light. This is perhaps the only possible
experimental support for the conjectural ``rolling-up'' of the wavelength
$\la_e$ around the sphere of radius $1l_e$ and
the wave equations (5), with $r_e=1l_e$ instead of $R=l_e/\al$.

The value of $\al$, which implies the reduction of scale between the world of
electromagnetic waves and the world of particles possessing mass, is due to
the interaction of the latter particles with zero-point radiation which,
thanks to the Compton effect, compresses them until equilibrium is reached
with elec\-tros\-ta\-tic repulsion, which in the case of the electron happens
when $r_e=1l_e$.

For the photon which has an energy equal to $m_ec^2$, we have $\la_e=2\pi
l_e/\al$, \ $R_e=l_e/\al$ \ and $r_e=1l_e$.
The decrease from $R_x$, which is a characte\-ris\-tic of the wavelength $2\pi
R_x$, to $r_x$, which is a characte\-ris\-tic of the particle, is always the
same; for any elementary particle with mass $m_x$, $r_x$ is always equal to
$\al R_x$.

The fine structure constant $\al$, the wavelength $q_\la$ of the most
energetic photon and the speed of the light, $c$, are three fundamental and
invariable physical entities which modulate the architecture of our Universe.

\section{Electrostatic and Centrifugal Forces  in the Suggested Configuration}

The suggested configuration implies that the whole of the substance of the
photon transformed into an electron maintains its coherence, and revolves
around the centre of the sphere of radius $l_e$, over whose surface it moves,
in such a way that the tangential velocity at all its points is equal to that
of the original photon, i.e. that of light.

The ``energy packet'' of dimensions $E=m_ec^2$, which in the configuration
described here revolves as a single entity around a point, from which all its
substance is distant by $l_e$, in such a way that its tangential velocity is
always $c$, implies a mass of $m_e=E/c^2$ and a centrifugal force of
$f_e=m_el_e^{-1}c^2=m_el_et_e^{-2}$. This force is exactly equal to that of
the elec\-tros\-ta\-tic repulsion of the electron's charge against itself on the
surface of the particle, i.e. to:
$$f_e=\que{e^2}{l_e^2}=\que{m_el_e^3}{t_e^2}\que1{l_e^2}=
\que{m_el_e}{t_e^2}=2.905352\ti10^6\ {\rm dyn}$$

The equivalence just stated leads us to suggest that the elec\-tros\-ta\-tic field
determined by the electron's charge, could proceed from the interaction
between the centrifugal force inherent to its spin
and zero-point radiation [2].

Figure 3 shows a section of the electron with centre at $O_1$, following
approximately a spire of the path given by equations (6).
The arrow $\overrightarrow{A_0T}$ shows the tangential velocity
``$c$'', constant along its length, while the other arrow
$\overrightarrow{A_0C}$ shows the
centrifugal force inherent to the mass of the electron, its radius and the
said tangential velocity.

\begin{figure}[h]
\centering
\resizebox{0.70\columnwidth}{!}{\includegraphics{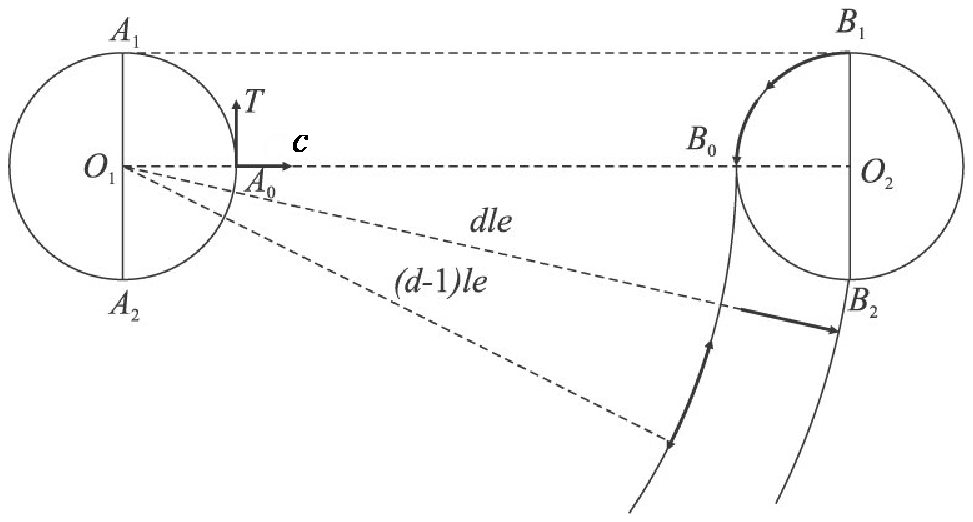}}
\caption{Fig. 3}
\end{figure}

If we imagine that the dimension of the radius of the electron extends as far
as the surface of the electron with centre at $O_2$, its centrifugal force at
the point of contact $B_0$ will be
$$f_c=\que{m_e}d\que1{dl_e} c^2=\que1{d^2}\que{m_el_e}{t_e^2}=
\que{e^2}{(dl_e)^2},$$
i.e. will be equal to the electrostatic repulsion between the two electrons.

When the particle with centre at $O_1$ and the particle with centre at $O_2$
have identical spins, the arrow which follows the circle with centre at $O_1$
and radius $dl_e$ faces at $B_0$ an opposed arrow following the circle with
centre at $O_2$ and radius $1l_e$. When their spins are opposed, those arrows
are in agreement. In the first case, the force $f_e=\que{e^2}{(dl_e)^2}$ is
one of repulsion, in the second one of attraction.

For this to be possible, the spin of the elementary particles must be always
around the same direction, which can be only the centre $O$ of the surface
$w^2+x^2+y^2+z^2=R^2$, which is the configuration of the Universe. This is not
impossible, as can be seen if we imagine minor circles on the surface
$x^2+y^2+z^2=R^2$, and assume that they spin around axes pointing towards the
centre of that surface.

In Fig. 3, we can also see how all this could be achieved through the
interactions of zero-point radiation and electrons. The photons which fall on
any point on the surface of the electron have three times the energy necessary
to balance the centrifugal energy generated by the rotation of the mass
$m_e/4\pi k^2_\la$ around $O$ with tangential velocity ``$c$'', see [2] pag.
10, and the sum of their projections over any given direction is equal to a
third of that energy. The radius of every electron measures $1l_e$, and the
line parallel to $\overline{O_1O_2}$ from $A_1$ on the surface of the electron
with centre at $O_1$ reaches a point $B_1$ on the surface of the electron with
centre at $O_2$.

The photons which reach $A_1$ disperse from that point in every direction,
after having generated the centripetal force required to balance the
centrifugal force at that point, and the intensity of the energy flows
emerging from $A_1$ is $\que1{d^2}3\va$, at the distance $d$ from $A_1$. The
sum of all the energy flows which emerge from the hemisphere of the electron
with centre at $O_1$ facing the electron with centre at $O_2$ is
$\que32\que{m_el_e^2}{t_e^2}$, for each $t_e$, at the distance $1l_e$ from
$O_1$ and $\que32\que1{d^2}\que{m_el_e^2}{t_e^2}$, for each $t_e$ at the
distance $dl_e$ from $O_1$. The sum of their projections over the direction
$\overline{O_1O_2}$ is $\que13\ti\que32\que1{d^2}\que{m_el_e^2}{t_e^2}$ for
each $t_e$, which generates a force $\que12\que1{d^2}\que{m_el_e}{t_e^2}$ in
this direction, when reaching the points $B$ at the surface of the electron
with centre at $O_2$. The presence of the electron with centre at $O_2$ has a
symmetrical effect on the electron with centre at $O_1$, whence the apparent
force of rejection $f_e=\que1{d^2}\que{m_el_e}{t_e^2}=\que{e^2}{(dl_e)^2}$.

\section{Suggestions on Elementary Particles of mass $m_x\not=m_e$}

The centrifugal forces, $f_{c1}$, inherent to the spin of the electron with
the configuration suggested in section 3 of this paper, are equal to the
centripetal forces, $f_0$, resulting from the interactions of the electron
with the zero-point radiation, whose intensity given by equation (15) in [2]
is independent from the characteristics of the said particle. In other words
$$f_{c1}=\que{m_e}{l_e}c^2=\que{m_el_e}{(t_e)^2}=f_0.$$

The quantic threshold $m_x(m_e) r_x(l_e)=m_el_e=e^2/c^2$, where ``$m_x$'' is
the measurement of the mass of the particle expressed in $m_e$, and ``$r_x$''
is that of the radius of the same particle expressed in $l_e$, allows us to
write:
$$f_{cx}=\que{m_x(m_e)}{r_x(l_e)}c^2=
\que{m_x(m_e) r_x(l_e)}{\{r_x(l_e)\}^2}c^2=
\que1{(r_x)^2}\que{m_el_e}{(t_e)^2}$$

For $m_x<m_e$, $r_x>l_e$, so that the total of the centrifugal forces will be
less than the total of the centripetal ones, and these centripetal forces
will squeeze the particle, whose radius will decrease to $r_y<r_x$. However,
its mass will remain the same, so that we have $m_xr_y<m_xr_x=e^2/c^2$, which
violates the said quantic threshold. Therefore, there cannot exist elementary
particles of spin $\hbar/2$ with a mass of less than 1~$m_e$, or with a radius
greater than 1~$l_e$.

The spherical surface of radius $r_e=1l_e$ is the geometrical location at
which there is produced an equilibrium between the centrifugal forces inherent
to elementary particles of spin $\hbar/2$, and the centripetal forces
resulting from the particles'\ interactions with zero-point radiation. Such an
equilibrium only happens for particles of mass $m_e$, i.e. electrons, and this
shows clearly their singular character as fundamental components of the
architecture of the Universe.

For $m_x>m_e$, the quantic threshold $m_x(m_e)r_x(l_e)=e^2/c^2$ prevents a
particle of mass $m_x$ from expanding beyond a spherical surface of radius
$r_x(l_e)=\que{m_e}{m_x}(l_e)$, and the centrifugal force determined by a mass $m_x(m_e)$
spinning with a radius of $r_xl_e$ and a tangential velocity ``$c$'' is:
$$f_{cx}=\que{m_x(m_e)}{r_x(l_e)}c^2=\que1{(r_x)^2}\que{m_el_e}{(t_e)^2}>
\que{m_el_e}{(t_e)^2}=f_0$$

For the proton, $m_x=(1.836153\ti10^3)m_e$; $r_x=(1.836153\ti10^3)^{-1}l_e$,
and therefore $f_{cx}=3.371457\ti10^6f_0$.

Faced with this centrifugal force, more than 3 million times greater than the
centripetal, a particle of mass $m_x>m_e$ could only reach equilibrium by
expanding its radius to measure $1.836153\ti10^3l_e$. This is impossible
because the quantic threshold $m_x(m_e)r_x(l_e)=e^2/c^2$ prevents it. The
enormous difference between $f_{cx}$ and $f_0$ at the limit imposed by this
quantic threshold gives rise to new quantic fluctuations of space, distinct
from those inherent to its curvature which are the cause of zero-point
radiation. These new fluctuations produce flows of photons in the direction of
the centre, $O_1$, of the particle in question, and their intensity is that
which is needed in order to balance out the centrifugal and the centripetal
forces, i.e. a force of:
$$f_{0x}=\left\{\que1{(r_x)^2}-1\right\}\que{m_el_e}{(t_e)^2},\eqno{(7)}$$
for the entire particle.

In the case of the proton, $f_{0x}=\{3.371457\ti10^6-1\}\que{m_el_e}{(t_e)^2}$
and we can assume that $f_{0x}=\que1{(r_x)^2}\que{m_el_e}{(t_e)^2}$, allowing
for a relative error $\vare<2.97\ti10^{-7}$. In the case of the electron
$f_{0x}=0$, and this does not allow for the appearance of any new quantic
fluctuations of space.

In [2] it was stated that, in the case of the electron, the intensity of the
energy flow per $(q_\la)^2$ is that which is required to generate the force:
$$f_{01}=\que1{4\pi(k_\la)^2}\que{m_el_e}{(t_e)^2}\qquad\hbox{per}\ \
(q_\la)^2;\qquad \hbox{see eq. (16) on [2].}$$

The mass $1m_e$ spinning at a tangential velocity ``$c$'' and with a radius
$1l_e$ causes a centrifugal force of $f_{c1}=\que{1m_e}{1l_e}c^2=
\que{m_el_e}{(t_e)^2}=f_0$. As remarked above this force is equal to the
elec\-tros\-ta\-tic repulsion against
itself of the electrical charge ``$e$'' confined
into a spherical surface of radius $l_e$.

We must here remember that the quantic threshold $m_x(m_e)r_x(l_e)=e^2/c^2$
does not only imply the spin $\hbar/2$, but also that the elec\-tros\-ta\-tic
repulsion $f_e=\que{e^2}{(r_xl_e)^2}=\que1{(r_x)^2}$ is equal to the
centrifugal force at $dl_e=r_xl_e$.

On the spherical surface of radius $r_e=1l_e$, i.e. the basic spherical
surface, the centrifugal force corresponding to the mass $m_x(m_e)$ spinning
at a tangential velocity ``$c$'' is:
$$f_{m_x1}=\que{m_x(m_e)}{1(l_e)}(c^2)=
\que1{r_x}\left(\que{m_el_e}{t_e^2}\right),$$
where the symbols into brackets mean units of the $(m_e,e,c)$ system.

This same mass, $m_x(m_e)$, when spinning at a tangential velocity ``$c$''
and radius of spin
$r_x(l_e)=\left(\que{m_e}{m_x}\right)l_e$,
gives rise to the centrifugal force:
$$f_{m_x,x}=\que{m_x(m_e)}{r_x(l_e)}(c^2)=
\que{m_x}{r_x}\left(\que{m_el_e}{t_e^2}\right)=
\que1{r_x^2}\left(\que{m_el_e}{t_e^2}\right),$$
because $m_xr_x=1$.

It therefore follows that between a point a distance of $1l_e$ from the centre
of the particle, and a point at a distance of
$r_x(l_e)<1l_e$ from that centre, the
centrifugal force of the elementary particles
is inversely proportional
to the distance to their centres, whilst the  intensities per $(q_\la)^2$ of
the energy flows around to balance the centrifugal and the centripetal forces
at a distance $r_x(l_e)$ from $O$, would concentrate from $\que{m_x}{4\pi
k_\la^2}\left(\que{m_el_e^2}{t_e^2}\right)$ each $t_e$, at the distance
$1l_e=k_\la(q_\la)$, to $\que{m_x}{4\pi k_\la^2
r_x^2}\left(\que{m_el_e^2}{t_e^2}\right)$ each $t_e$, at the distance
$r_x(l_e)=k_\la r_x(q_\la)$ from $O$, which means intensities inversely
proportional to the squares of the distances. In the following lines we will
suggest a hypothesis to explain this.\vskip 1pc

\begin{center}*\ \ \ *\ \ \ *\end{center}\vskip 16pt

We have supposed that the flows of photons which arise to balance the
centrifugal force on the surface of nucleons, are directed towards the centres
of these particles. In this respect, they act differently from zero-point
radiation, which is made up of photons which can follow any trajectory.

The radius of the electron is extremely small, $r_e=2.817940\ti10^{-13}$ cm,
but the elementary areas $(q_\la)^2$ which compose its surface are even
smaller. They number $8.333333\ti10^{-42}$ and the distance between the
centres of contiguous areas is $1q_\la=3.460408\ti10^{-34}{\rm
cm}=(2\al\pi)^{1/2}L_P$, where $L_P$ is Planck's Length.

We have no experience of the behaviour of the photons flows which, separated
by distances of this order of magnitude, are almost in parallel, but we can
suppose that they can interreact with each other, with the photons behaving as
particles, instead of doing so through the interference of their respective
waves. If so, the number of collisions per unit of time will be directly
proportional to the densities of the flows, and inversely proportional to the
squares of the distances to the centre towards which they converge.

It is logical to suppose that these collisions between particles will produce
changes of direction, which will surely begin at a distance from the centre
which is much greater than the radius of the electron. Initially, the
frequency of these collisions will be insignificant.

We know that the intensity of the flows of photons which converge towards the
centre of the proton is, at a distance $l_p$ from that centre, that which is
needed for their interactions with the particle to produce a centripetal force
per $(q_\la)^2$ which equals the centrifugal force generated by the particle's
spin, of radius $r_p$ and tangential velocity $c$; in other words, equal to
$m_p(r_p)^{-1}l_e^{-1}c^2=m_p(r_p)^{-1}l_et_e^{-2}$, where $m_p$ is the mass
of the proton, and $(r_p)=m_em_p^{-1}$, the number which results from
measuring the radius of the proton in units of length $l_e$.

If the intensities of the photon flows diminish, starting at a distance $r_p$
from the centre of the proton, $O$, in inverse proportion to the square of the
distance to it, we obtain, for the distance $r_x>r_p$:
\begin{eqnarray*}
\va_x & = & \va_p\left\{1-\Bigl(\que{dx}{(r_p)^2}+\que{dx}{(r_p+1)^2}+\cdots+
\que{dx}{x^2}\Bigr)\right\}\\
& = & \va_p\left(1-\int^{r_x}_{r_p}\que{dx}{x^2}\right).
\end{eqnarray*}

For $x=1l_e$, we obtain:
$$\va_{l_e}=\va_p\left(1-1+\que1{r_p}\right)=\que{\va_p}{r_p}.$$
However, this intensity has been dispersed over the $4\pi l_e^2(q_\la)^2$ of
the surface of the electron, from the $4\pi(r_p)^2l_e^2(q_\la)^2$ of the
surface of the proton, so that the intensity of the flow on the surface of the
electron must be multiplied by $\left(\que{r_p}{l_e}\right)^2$, with the
result that at a distance of $l_e$ from $O$, the flow is of $\va_pr_p$, which
could cause a force equal and opposite to that of the mass of the proton
rotating with a radius of $l_e$ and a tangential velocity of $c$, since
$$\que{m_p}{(r_p)l_e}\c(r_p)c^2=\que{m_p}{l_e}c^2=\que{m_pl_e}{t_e^2}.$$

The photons which have been dispersed from their trajectories towards ``$O$"\
continue along different ones, but all the photons which have entered the
spherical surface of radius $r_e$ will exit from it, and the sum of the
possible exchanges of energy between them will be nil, thus producing an exact
equality between the energies which enter and exit.

For any distance $r_p<x<l_e$, we can write:
$$\va_{l_e}= \va_x \left(1-\int^{l_e}_x \que{dx}{x^2}\right)= \va_x  \que{1}{(x)}.$$
For a distance $l_e$ from $O$, we must multiply by $\left(\que
x{l_e}\right)^2$, from which results $\va_{l_e}=\va_{x}\que1{(x)}\left(\que
x{l_e}\right)^2=\va_x  x$, whence $\va_x=\va_{l_e} \que1x$; that is, equal to
$\va_{l_e}$ multiplied by the inverse of the relation of distance from ``$O$".

\begin{center}*\ \ \ *\ \ \ *\end{center}\vskip 16pt

\vskip 6pt
\begin{center}*\ \ \ *\ \ \ *\end{center}\vskip 16pt

Figure 4 shows two elementary  particles which are $dk_\la(q_\la)$ apart, the
first having mass $m_x(m_e)$ and centre at $O_1$, and the second having mass
$m_y(m_e)$ and centre at $O_2$.

The energy flow aroused by the inferiority of the centripetal forces to the
centrifugal ones on the surface of the particle with centre at $O_1$ and mass
$m_x$, has at $A_2$, which is $1l_e$ distant from $O_1$, an intensity of:
$$\va_{1x}=\que{m_x}{4\pi k_\la^2}\left(\que{m_el_e^2}{t_e^2}\right)\qquad
\hbox{per}\ (q_\la)^2\ \ \hbox{each}\ t_e.\eqno{(8)}$$
At  a distance of $d(l_e)=k_\la d(q_\la)$ from $O_1$, this flow, which
proceeds from the $4\pi k_\la^2$ areas $(q_\la)^2$ which are $k_\la(q_\la)$
distant from $O_1$, is spread over $4\pi k_\la^2d^2(q_\la)^2$, with a
resulting diffusion of $4\pi k_\la^2/4\pi k_\la^2d^2=1/d^2$.

\begin{figure}[h]
\centering
\resizebox{0.70\columnwidth}{!}{\includegraphics{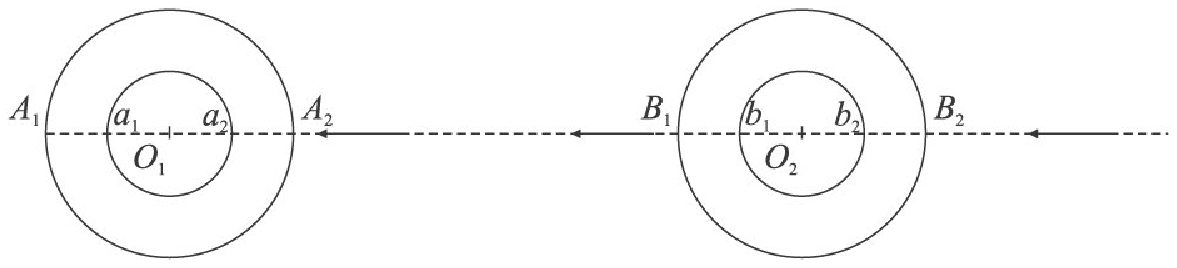}}
\caption{Fig. 4}
\end{figure}

The lessening of the intensity of the flows $\va_{1x}$ which flow in from each
of the $4\pi k_\la^2$ areas of $1(q_\la)^2$ on the spherical surface of radius
$k_\la(q_\la)$, is not due to a spatial diffusion of the demand from the area
$4\pi k_\la^2(q_\la)^2$ to the area $4\pi k_\la^2 d^2q_\la^2$; each of these
maintains its identity from one of the $4\pi k_\la^2 d^2$ areas $(q_\la)^2$,
which are $k_\la d(q_\la)$ distant from $O_1$, up to the $4\pi k_\la^2$ areas
$(q_\la)^2$ on the said spherical surface. What is produced is a temporal
diffusion. During the lapse of time ``$t_e$'' the flows onto each $(q_\la)^2$
on the spherical surface of radius $k_\la(q_\la)$ proceed from any one of
those which belong to the spherical segment of one base which is determined by
the intersection of the cone with vertex at $O_1$ and base on that
$(q_\la)^2$, with the spherical surface of radius $dk_\la(q_\la)$ and centre
at $O_1$. It may be any one of those which exist at the said intersection, and
the probability of being any particular one during the lapse $1(t_e)$ is
$1/d^2$. Therefore, the intensity of the flow which arrives from any one of
them at one of those situated at a distance of $k_\la(q_\la)$ from $O_1$ is:
$$\va_{1d}=\que1{4\pi
k_\la^2}\que{m_x}{d^2}\left(\que{m_el_e^2}{t_e^2}\right)\qquad\hbox{per}\ \
(q_\la)^2\qquad\hbox{each}\ \ t_e.$$

The flow $\va_{1x}=\que{m_x}{4\pi k_\la^2}\left(\que{m_el_e^2}{t_e^2}\right)$
per $(q_\la)^2$ each $t_e$, aroused at the area $2\pi k_\la(q_\la)^2$ of the
spherical segment of one base, height $1(q_\la)$ and centre at $A_2$, falls
every $d^2t_e$ on the analogous spherical segment with centre at $B_2$, on the
spherical surface with centre at $O_2$ and radius $k_\la(q_\la)$. From here,
and in response to the demand provoked by the inferiority of the centripetal
forces to the centrifugal ones on the surface of the elementary particle of
mass $m_y$ and centre on $O_2$, it converges towards $O_2$, suffering
interferences which causes its intensity per $(q_\la)^2$ at $b_2$ to be:
$$\va_{m_x,m_y}=\que1{4\pi
k_\la^2}\que{m_x}{r_y}\left(\que{m_el_e}{t_e^2}\right)\ \ {\rm per}\ \
(q_\la)^2\ \ {\rm each}\ \ d^2t_e,$$
and it causes there to be, on the $(q_\la)^2$ situated at $b_2$ a force along
$\ovl{O_1O_2}$ with an intensity of:
$$f_{(m_x,m_y,d)}=\que1{4\pi
k_\la^2}\que{m_xm_y}{d^2}\left(\que{m_el_e}{t_e^2}\right)$$

The presence at $O_1$ of a particle of mass $m_x$ determines that there exists
at $a_1$, on $\ovl{A_1A_2}$ and at a distance $r_x$ from $O_1$, a force of
equal intensity directed towards $O_2$, so that there results an apparent
attraction between the two particles given by:
$$f_g=\que{m_xm_y}{2\pi
k_\la^2}\que1{d^2}\left(\que{m_el_e}{t_e^2}\right).$$

In the $(e,m_e,c)$ system, the value of the numerical coefficient of the
gravitational constant is $G_e=1/2\pi k_\la^2$. Therefore the force $f_g$ is
equal to the gravitational attraction between two particles which are $dl_e$
apart and whose masses are $m_x$ and $m_y$ respectively.

\vskip 6pt
\begin{center}*\ \ \ *\ \ \ *\end{center}\vskip 16pt

The photon flows of zero-point radiation
which fall incessantly on every $(q_\la)^2$ of the surface of
an elementary particle of mass $m_x$ and radius $r_x=\que{m_e}{m_x}l_e$, with
the intensity needed to generate a force equal and opposite to the centrifugal
force:
$$f_e^x=\que{m_x(m_e)}{r_x(l_e)}c^2=
\que1{(r_x)^2}\que{m_el_e}{(t_e)^2}=\que{e^2}{(r_xl_e)^2}$$
must necessarily oppose that force at all times. The spin $\hbar/2$ is a
quantic threshold; the flows cannot be converted into mass added to that of
the particle, and since they cannot accumulate in it
they must pass out of it. They
can do this along a tangent to the line which follows the photon rolled up
around the spherical surface of radius $r_x$ (see the hypothesis suggested in
Section 3 of this paper) producing a configuration of photons with spheres of
increasing radii, and flows whose intensities decrease in inverse proportion
to the squares of those radii. Their intensity at the spherical surface of
radius $r_x$ is that which is needed to equal the elec\-tros\-ta\-tic repulsion
against itself of the charge ``$e$", distributed evenly over that surface, in
other words, to equal the force
$\que{e^2}{(r_xl_e)^2}=\que{m_x(m_e)}{r_x(l_e)}c^2.$
At a distance of $1l_e$ from $O_1$, the intensity $\va\pri_0$ is that which is
needed to equal the force $f\pri_0=\que{e^2}{l_e^2}=\que{m_el_e}{t_e^2}$, and
at a distance $d$, the intensity $\va^d_0$ is that which is needed to equal
the force
$$f^d_0=\que{e^2}{(dl_e)^2}=\que1{d^2}\que{m_el_e}{t_e^2}$$

The revolutions along the said spheres are in the direction in which the
photon is rolled up into the particle. If they meet with another particle
which has the photon rolled up in the same direction, they will go in the
opposite direction to it (see Fig. 5) and, since their uninterrupted flows
cannot have any effect on the particle in question, they must exit from it. And
if, previously, flows directed towards the centre of the particle must produce
exits along tangents to the lines of rolling-up, the flows which follow the
tangents will now give rise to a single flow towards $O_1$, having an
intensity of $\que1{d^2}\que{m_el_e}{t_e^2}$, where $d(l_e)$ is the distance
$\ovl{O_1O_2}$. Since we are now dealing with all the flows which strike upon
the hemisphere of the particle or which are emitted from it, the value of the
force of repulsion directed along $\ovl{O_1O_2}$ is:
$$\que{2\pi(k_\la)^2}{4\pi(k_\la)^2}\que1{d^2}\que{m_el_e}{t_e^2}=
\que12\que1{d^2}\que{m_el_e}{t_e^2}$$
and, since the presence of the particle at $O_2$ has the same effect on the
particle whose centre is at $O_1$, there results an apparent repulsion between
the two, which  is equal to the quotient $\que{e^2}{d^2l_2^2}$, which is $2\pi
k_\la^2$ times more intense than gravitational attraction.

\begin{figure}[h]
\centering
\resizebox{0.70\columnwidth}{!}{\includegraphics{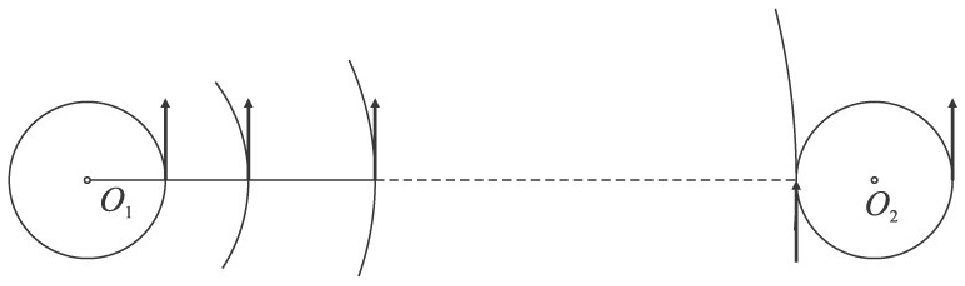}}
\caption{Fig. 5}
\end{figure}

If the particle at $O_2$ is configured as a photon rolled up in the direction
opposite to that of the particle whose centre is at $O_1$, there will be
produced an attraction instead of a repulsion.

The hypothesis set out in Section 3 of this paper enables us to explain how a
spin of $\hbar/2$ can be produced in the elementary particles. Through the
help of that hypothesis and suggestions made in the present section, we can
also explain the causes of elec\-tros\-ta\-tic attraction and repulsion and of the
existence of positive and negative electric charges, and how gravitational
attraction and the elec\-tros\-ta\-tic forces are due to one and the same cause;
this compels the intensity of the elec\-tros\-ta\-tic forces to be $2\pi(k_\la)^2$
times that of gravitational attraction, if
$k_\la=\left(\que1{2\pi\al}\right)^{1/2}\que{l_e}{L_P}$
where $l_e$ is the ``classical radius"\ of the electron, and $L_P$ the Planck
length. This simple relation between the gravitational constant,
$G_e=\que1{2\pi(k_\la)^2},$ the fine structure constant $\al$, the ``classical
radius"\ of the electron, $1l_e$, and the Planck length
$L_P=\left(\que1{2\pi\al}\right)^{1/2}q_\la$, can be admired as a fascinating
harmony within the world of the elementary particles and fundamental forces.

\section{The Energy Flows resulting from the presence of particles of mass
$m_x>m_e$, and the Cohesion of Atomic Nuclei}

The title of this section does not mean that we mean to propose a model for
atomic nuclei, but that, much less ambitiously, we want to show that the
suggestions made in the previous sections lead to the conclusion that if they
fit with physical reality, the energy flows resulting from the presence of the
protons and neutrons which make up atomic nuclei, would possess the intensity
needed to ensure the cohesion of those nuclei.

The reasoning developed in order to reach this aim rests on a basis
consisting of the said suggestions, through the following premises, taken from
[4], pp. 48-50:
\begin{enumerate}\itemsep=0pt
\item The number of nucleons, $A$, per unit of volume is roughly constant,
$\que A{(4\pi/3)R^3}\simeq$ constant, whence $R=R_0A^{1/3}$, where $R$ is the
radius of the nucleus.
\item From electron scattering measurements it is concluded that:
$$R_0=1.2fm=0.4258l_e$$
\item The density charge changes very little from the lightest nuclei to the
heaviest up to a certain point, and then declines slowly over a certain
distance, over which the charge falls from 90\% to 10\% of its central value.
The length of this distance is approximately $2.3fm=0.8162le$.
\item For light elements, the number of protons $N_P$ is approximately equal
to the number of neutrons $N_n$. For heavy elements, $N_P\simeq0.41A$ (See
[4], p. 70).
\item The binding force for each nucleon is roughly equal to $8M_eV$ (See [4],
p.~67).
\item The nuclear force must be spin-dependent (See [4], p. 92).
\item In [4], Fig. 3.4 on p. 49, we can see the curves which express the
radial distribution of the charges for certain nuclei. The lightest is that of
C$_{12}$, with 6 protons and 6 neutrons $(A=12,\ N_p=N_n)$, and the heaviest
is that of Pb$_{208}$, with 82 protons and 126 neutrons $(A=208,$ $N_p=82,$
$N_n=126,$ $N_p=0.394A)$.

In the nucleus of C$_{12}$, the radius of the nucleus measures $0.9748l_e$,
and 90\% of its charge is within a sphere with its centre at the centre of the
particle, and with a radius of $(0.9748-0.8162)l_e=0.1586l_e$. In the case of
Pb$_{208}$, the radius of the nucleus measures $2.5228l_e$, and 90\% of the
particle lies within a sphere with the same centre and a radius of
$1.7067l_e$.
\end{enumerate}

Figure 6 presents the case of two elementary particles, one of them with centre
at $O_1$, a mass of $m_x(m_e)$, and a radius of $r_x=\que{m_e}{m_x}l_e$. The
other has its centre at $O_2$, at a distance of $dl_e<l_e$ from $O_1$, a mass
of $m_y(m_e)$, and a radius of $r_y(l_e)=\que{m_e}{m_y}l_e$.

\begin{figure}[h]
\centering
\resizebox{0.70\columnwidth}{!}{\includegraphics{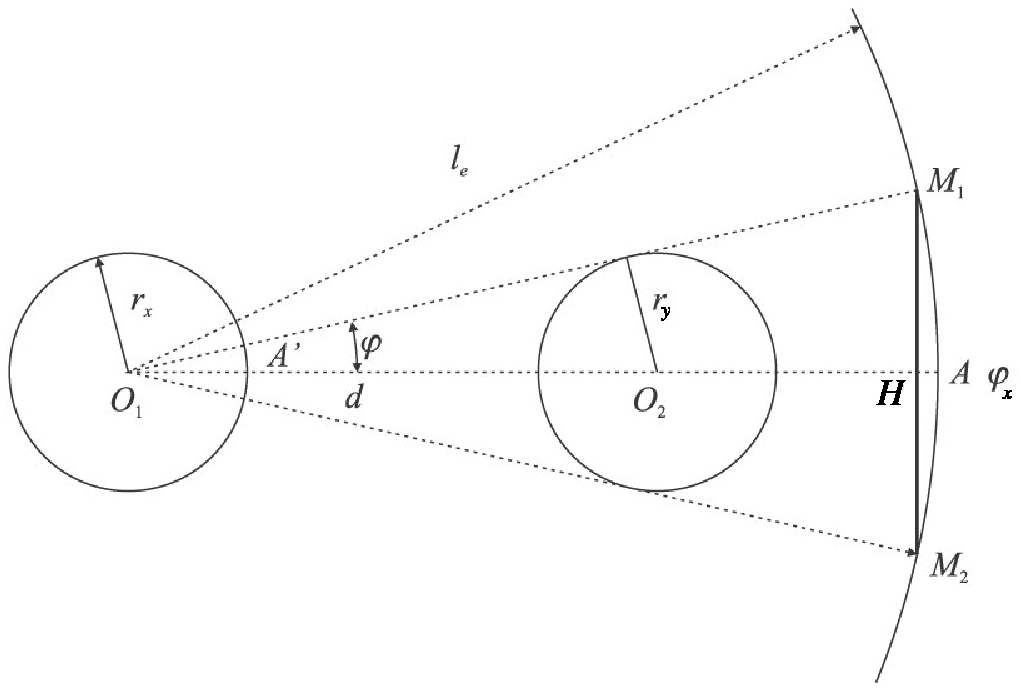}}
\caption{Fig. 6}
\end{figure}

In this figure, the symbol $\va_x$ signifies the intensity at $A$ of the
energy flow required in order to generate the force
$f_{(x,1)}=\que{m_x}{4\pi(k_\la)^2}\left(\que{m_el_e}{(t_e)^2}\right)$ per
$(q_\la)^2$. The presence of the particle with centre at $O_2$ and radius
$r_y$ prevents the flows aroused in the spherical segment of one base
$\widehat{\overline{M_1AM_2}}$,
on the spherical surface of radius $1(l_e)$,
from arriving at the particle with centre at $O_1$. The area of this segment
is:
$$S_{1x}(q_\la)^2=2\pi k_\la(q_\la)\overline{AH}(q_\la)=
2\pi k_\la\ovl{AH}(q_\la)^2;$$
$$\ovl{AH}=l_e(1-\cos\va),$$
where $\sin\va=r_y/\ovl{O_1O_2}=r_y/d$ whence
$$S_{1x}(q_\la)^2=2\pi(k_\la)^2\Bigl[1-\{1-(r_y/d)^2\}^{1/2}\Bigr](q_\la)^2.$$

The blocking of the flows $\va_0^\prime$, aroused in the areas
$S_{1x}(q_\la)^2$ of the spherical segment of one base
$\widehat{\overline{AM_1M_2}}$, prevents them from becoming
concentrated on the spherical segment which corresponds to the spherical
surface of radius $r_x$ and centre at $O_1$, whose surface is
$$S_{xx}(q_\la)^2=2\pi(k_\la)^2(r_x)^2
\Bigl[1-\{1-(r_y/d)^2\}^{1/2}\Bigr](q_\la)^2.$$
Such a concentration would cause their intensity per $(q_\la)^2$ on the
surface of radius $r_x$ and centre at $O_1$ to be $(1/r_x)^2$ times
$\va_{x1}$; however, the interference between flows, as explained above,
causes that intensity per $(q_\la)^2$ to be only $(1/r_x)\va_{x1}$, which
implies that the total of the energy flows which do not arrive from the
surface of the spherical segment of one base which has an area of
$S_{1x}(q_\la)^2$, to the spherical segment which has an area of
$S_{xx}(q_\la)^2$, is that required to generate a force of:
$$FS_{xx}=2\pi(k_\la r_x)^2\left[1-\Bigl(1-\{1-(r_y/d)^2\}^{1/2}\Bigr)\right]
\que{m_x}{m_y}\left(\que{m_el_e}{t_e^2}\right)$$

The electrostatic repulsion at a distance $r_x q_\la$ from $O_1$ determines a
bundle of forces equal and opposed to the forces $FS_{xx}$, and when they are
not counteracted by the latter forces, they move towards the exterior. The sum
of the projections of these forces along the line $\ovl{O_1O_2}$ determines a
force along $\overrightarrow{O_1O_2}$.

Figure 7 shows the generation of the spherical surface with centre at $O_1$ and
radius $r_x(l_e)$, through differential surface elements, consisting of rings
of radius $r_x(l_e)\sin\va$ and width $r_x(l_e)d\va$ orientated along the
tangent at $B$. In effect,
$$dS=2\pi r_x(l_e)\sin\va r_xl_ed\va=2\pi(r_x)^2l_e^2\sin\va,\eqno{(11)}$$
whence $S=(1/2)$ the area of the spherical surface of radius $r_xl_e$ is:
$$S=2\pi(r_xl_e)^2\int^{\pi/2}_0\sin\va d\va=2\pi(r_x)^2(l_e)^2$$

The force of intensity $f_{xx}$, directed along the radius $\ovl{OB}$, can be
broken down into two: $f\cos\va$, its projection along the radius $\ovl{OA}$,
and $f\sin\va$, which is cancelled out when added to the equal and opposing
force which corresponds to $f_{xx}^\prime$ along the radius $\ovl{OB_1}$,
which is symmetrical in $\ovl{OA}$ with respect to $\ovl{OB}$. This leads to
the result: $dF=f_{xx}dS$; whence
$dF\cos\va=f_{xx} 2\pi(r_xl_e)^2\sin\va\cos\va
d\va$, and:
$$\begin{array}{c}
\disp F=f_{xx}\pi(r_xl_e)^2\int^\va_02\sin\va\cos\va d\va=
f_{xx}\pi(r_xl_e)^2\Bigl[\sin^2\va\Bigr]^\va_0 \\ \\[-4pt]
\disp=f_{xx}\pi(r_xl_e)^2\sin^2\va,\end{array}\eqno{(12)}$$

Fig. 8 shows that $\sin\va=r_y/\ovl{O_1O_2}=r_y/d$, and also
$$f_{xx}=\que1{4\pi(k_\la r_x)^2}
\que{m_x}{r_x}\left(\que{m_el_e}{t_e^2}\right)=
\que1{4\pi(k_\la r_x)^2}
\que1{(r_x)^2}\left(\que{m_el_e}{t_e^2}\right).$$
By introducing these values into (12), we obtain:
$$F=\que1{4\pi(k_\la r_x)^2}\que1{(r_x)^2}
\left(\que{m_el_e}{t_e^2}\right)\pi(r_x l_e)^2\left(\que{r_y}d\right)^2.$$
Remembering that $1l_e=k_\la(q_\la)$, we finally arrive at:
$$F=\que14\left(\que{r_y}{r_x}\right)^2\que1{d^2}
\left(\que{m_el_e}{r_e^2}\right)$$
where $r_x=$ radius of the proton, and $r_y=$ radius of the neutron,
$\que{r_x}{r_y}=0.99725$; $\que{r_y}{r_x}=1.002757$ and when both particles
are equal $\que{r_x}{r_y}=1$. When we consider the reciprocal effect of the
particle with centre at $O_1$, on the particle with centre at $O_2$, and the
sum of the two, we obtain:

\begin{figure}[h]
\centering
\resizebox{0.70\columnwidth}{!}{\includegraphics{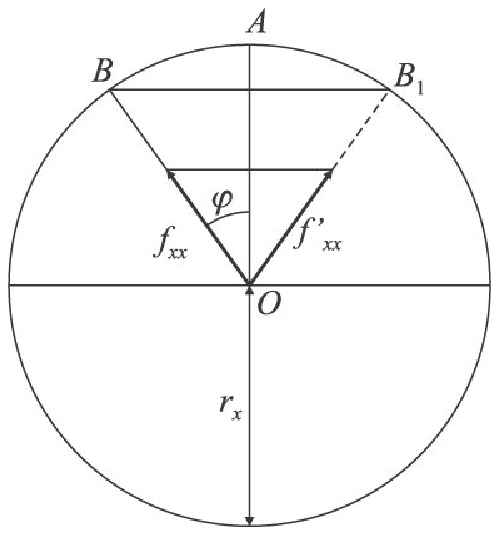}}
\caption{Fig. 7}
\end{figure}

For the pairs ``proton-proton"\ and ``neutron-neutron":
$$F=\que12\que1{d^2}\left(\que{m_el_e}{t_e^2}\right).$$

For the pairs ``proton-neutron":
\begin{eqnarray}
F  & = & \que{0.99725+1.002757}4\que1{d^2}\left(\que{m_el_e}{t_e^2}\right)\nonumber \\
& = & \que{2.0000076}4\que1{d^2}\left(\que{m_el_e}{t_e^2}\right)\simeq
\que1{d^2}\left(\que{m_el_e}{t_e^2}\right); \nonumber \\
\vare & < & 3.8\ti10^{-6} \nonumber
\end{eqnarray}
For heavy nuclei, the number of protons $N_p$ is approximately equal to
$0.41A$. We must therefore combine the elec\-tros\-ta\-tic force of repulsion
0.41
times with that of gravitational attraction, i.e. a mutual attraction of
$f_g=\que{0.5}{d^2}\que{m_el_e}{(t_e)^2}$, with a mutual repulsion of
$0.41f_e=\que{0.41}{d^2}\que{m_el_e}{t_e^2}$; this implies a slight predominance
of the centripetal forces, which can be compensated by the centrifugal forces
of the orbits of the nucleons within the atomic nucleus. This complies with the
proposition initially stated, which was to show that the energy flows which
arise to balance out the centripetal and centrifugal forces in elementary
particles of mass $m_x>m_e$ have sufficient intensity to generate the strong
interaction needed to ensure the cohesion of atomic nuclei.

\vskip 8pt
\begin{center}*\ \ \ *\ \ \ *\end{center}\vskip 12pt

It will be useful to consider another method of showing that the energy flows
which arise to balance out the enormous differences between the centripetal
and centrifugal forces on the surface of the nucleons, are sufficient to
generate a gravitational attraction with an intensity analogous to the
elec\-tros\-ta\-tic repulsion between two protons whose nuclei are separated by a
distance of less than $1l_e$. To make a first approximation, we can firstly
consider the mean distance between the particles. Premises 1, 2 and 3 taken
from [4] lead to the equation:
$$A(4\pi/3)(d_ml_e)^3=(4\pi/3)(0.4258A^{1/3}l_e)^3,\eqno{(13)}$$
whence $d_m=0.4258l_e$.

The number of protons is approximately equal to $0.4A$ for heavy nuclei, where
$A$ is the number of nucleons, as it is said in premise 7 taken from [4].

Within a sphere of radius $1l_e$ around the centre, $O_1$, of a proton, the
intensities of gravitational attraction decrease in inverse proportion to the
distance from $O_1$,
but the intensities of electrostatic repulsion decrease in inverse
proportion to
the square of this distance. A proton of mass $m_p$ and radius
$r_p=(m_e/m_p)l_e$ possesses $4\pi(k_\la)^2(r_p)^2$ areas $(q_\la)^2$ on its
surface. All of them play a part in electrostatic repulsion, but on
the gravitational attraction between the protons with centre at $O_1$ and
another nucleon with centre at $O_2$, less than $1(l_e)$ away from $O_1$, the
only influence is that exerted by the energy flows intercepted by the nucleon
with centre at $O_2$ and direction $\ovl{O_1O_2}$.

The centripetal force corresponding to the mass $m_p$ spinning at a tangential
velocity of ``$c$"\ and a radius of spin of $1(l_e)$ is $m_p(m_ec^2)/1(l_e)=
m_p\left(\que{m_el_e}{t_e^2}\right),$ and the centrifugal force per unit of
surface is:
$$f_{r_p}=\que1{(r_p)^2}\que{m_p}{r_p}\left(\que{m_el_e}{t_e^2}\right)=
m^4_p\left(\que{m_el_e}{t_e^2}\right),\eqno{(14)}$$
because $m_pr_p=1$.

The energy flows which could cause gravitational attraction between the
particle of mass $m_p$, radius $r_p$ and centre at $O_1$ and the similar
particle with centre at $O_2$, are those which, being directed towards $O_1$
fall on the spherical segment of one base, which belongs to the sphere with
centre at $O_1$ and radius $1(l_e)$, whose height is $h=1(q_\la)$ and which
has its centre at $A$. The area of this spherical segment is $S_1=2\pi
k_\la(q_\la)^2$, and the projection of this area over the plane perpendicular
to $\ovl{O_1A}$ at $O_1$ does not differ significantly from $S_s$. The sum of
the flows which fall on the half spherical surface with centre at $O_1$ and
radius $r_pl_e$ is proportional to $2\pi(k_\la r_p)^2(q_\la)^2$ and the
projection of this surface over the said plane perpendicular to $\ovl{O_1A}$
at $O_1$ has the area $S_{r_p}=\pi(k_\la r_p)^2(q_\la)^2$.

The quotient $\que{S_s}{S_{r_p}}=\que2{k_\la(r_p)^2}$ is the factor of area,
$\va_a$, by we must multiply the force per unit of surface at the distance
$r_p(l_e)$ from $O_1$, in order to obtain half the intensity of the energy
flow which generates the gravitational force. This flow arrives at a distance
$dl_e$ from $O_1$ multiplied by the factor $1/d$, because up to distance
$1(l_e)$ from $O_1$ the said flow  decrease in intensity in inverse ratio to
the distances from $O_1$, and not to the square of those distances. Taking all
of this together, we obtain $f_G/2=\va_a f_{rp} l_e/dl_e$. The presence of
the particle with center at $O_1$ exerts an equal and opposite effect on the
particle with center at $O_2$, so that we finally obtain:
$$f_G=\que{2\va_af_{rp}}d=\que4{k_\la(r_p)^2}
\que{(m_p)^4}d\left(\que{m_el_e}{t_e^2}\right),\qquad\hbox{i.e.}$$
$$f_G=\que{4(m_p)^6}{k_\la d}\left(\que{m_el_e}{t_e^2}\right),\qquad
\hbox{whence}\eqno{(15)}$$
$$m_p=\left(\que{0.1k_\la}{0.4258}\right)^{1/6}=2400,\qquad\hbox{i.e.}\quad
m_p=2400 m_e$$

Trying for a better approximation we can write:
$$\que{4(m_p)^6}{k_\la}
\left(\que1{d_1}+\que1{d_2}+\cdots+\que1{d_n}\right) $$
$$ =
0.4\left [ \que1{(d_1)^2}+ \que1{(d_2)^2}+\cdots + \que1{(d_n)^2} \right ],$$
in which $d_1,d_2,\l ,d_n$, are the distances between a proton and the
nucleons which are at a distance of less than $1l_e$ from its centre. Premise
3 in [4] implies a uniform distribution of charges in the heart of the
nucleus, and a mean distance of $0.4258l_e$ determines that in such a case
there are within a sphere of radius $1l_e$ around the proton:
$$n=(0.4258)^{-3}\ \ \hbox{nucleons, i.e., almost exactly 13 nucleons}$$

For $n=13$, \ $\disp\sum^n_1n^{-1}=3.180134$; \ \ $\disp\sum^n_1n^{-2}=1.570894$ \
and we obtain
$$\que{4(m_p)^6}{k_\la}3.180134=0.41.570894m_e,$$
whence:
$$m_p=\left(\que{k_\la0.4 1.570894}{12.720583}\right)^{1/6}m_e=
1.851m_e,$$
which is a very good approximation to the true value $m_p=1836m_e$

\section*{References}

[1] R. Eisberg y R. Resmic: {\em F\'{\i}sica Cu\'{a}ntica}. Ed. Limus S.A.
1983, pp. 326-328.

[2] R. Alvargonz\'alez: {\em Interactions bet\-ween ze\-ro-point ra\-dia\-tion
an\-d elec\-trons: arXiv physics}/0311139, V2, 23 May 2005.

[3] R. Feynman: {\em The Feynman Lectures in Physics}. Addison Wesley, 1964.

[4] K. S. Krane: {\em Introductory Nuclear Physics}. John Wiley \& Sons, 1988.
\end{document}